\documentclass[manuscript]{aastex}

\begin{document}

\title{A Peculiar Linear Radio Feature In The Supernova Remnant N206} 

\author{Robert J.  Klinger, John R. Dickel, Brian D. Fields}
\affil{Astronomy Department, University Of Illinois at Urbana-Champaign, 
    Urbana, IL 61801}

\and

\author{Douglas K. Milne}
\affil{Australia Telescope National Facility, Epping NSW 1710, Australia}

\begin{abstract}

We present images of the supernova remnant N206 in the LMC, taken with ATCA
at wavelengths of 3 and 6 cm.  Based on our data and previously published 
flux densities,
the spectral index of N206 is $-$0.20 $\pm$ 0.07.  The 6-cm radio morphology
shows a filled center.  Most interesting is the
discovery of a peculiar linear feature previously undetected at any
wavelength.  The feature lies to the east of the center of the remnant,
stretching from about one-fourth to three-fourths of the remnant's radius.  It
is wedge-shaped, with a steady opening angle from an apex on the eastern side.
The feature resembles
the disturbance expected from an object moving through the material
supersonically at about 800 km/s.  We present arguments suggesting
that the linear feature might have been produced by a low-mass star or compact
object ejected from a binary system which may have 
led to a Type Ic supernova.

\end{abstract}


\section{Introduction}

Several studies of supernova remnants (SNRs) in the
Large Magellanic Cloud (LMC) have included the SNR located
on the north-eastern edge of the \ion{H}{2} region LH$\alpha$ 120-N206 
(Henize 1956).   In following common practice, we will henceforth use the name
N206 for the SNR or use its coordinate designation B0532-710. 

Radio observations were made at several frequencies, ranging from
0.4 to 14.7 GHz (Mathewson \& Clarke 1973;
Milne, Caswell, \& Haynes 1980;
Milne et al 1984) with resolutions on the order of a few
arcminutes.  The data showed a bright source with a relatively flat radio
spectrum with an index of about -0.33
(Milne, Caswell, \& Haynes 1980).  However, the observations were
complicated by the presence of the nearby \ion{H}{2} region to the south-west 
of the SNR and were unable to resolve any structure in the interior
of the remnant.

H$\alpha$ images of N206 show a roughly circular filamentary shell 
around the periphery (Williams et al 1999).  Images
taken in [\ion{S}{2}] and [\ion{O}{3}] (Lasker 1977) also show a 
shell structure with little
central emission.  When combined with the earlier radio data, the optical 
images indicate that the object could be considered a shell type SNR.

X-ray data have been obtained with both Einstein and {\it ROSAT}.  
These images show
a centrally brightened morphology with no apparent edge brightening to
denote an X-ray shell (Long, Hellfand, \& Grabelsky 1981; Williams et al
1999).  By X-ray morphology alone, the remnant is classified as
centrally-brightened.  However, when combined with the radio and optical
morphology, the remnant has been suggested to be of the class
called ``mixed morphology'' (Rho \& Petre 1998), due to the disparity 
between the observed morphologies when viewed at different wavelengths.  
The centrally concentrated X-ray emission from mixed morphology remnants is 
thermal.  Currently available  
X-ray data are insufficient to determine whether or not the 
emission is thermal.

In this paper, we present high resolution (Half-Power Beam Widths (HPBW) 
$1.1''$ and  $1.8''$) radio images
of N206 at 3 and 6 cm made with the Australia Telescope Compact Array (ATCA).
These images show no detailed structure in the remnant
except for a linear feature that was
previously undetected at radio wavelengths and is also absent from both
optical and X-ray images of the SNR.

The observations and data reduction are discussed in Section 2 of this
paper.  Radio images and descriptions of the radio morphology, spectral
information, and polarization measurements are presented in Section 3.  
A discussion of the observed features is in Section 4.  Conclusions are
in Section 5. 

\section{Observations}

We observed N206 with the Australia Telescope Compact Array (ATCA).  Data
were taken on five observing nights in late 1997.  The observing parameters  
and observing dates are given in Table 1.  The baselines
were distributed between 31 m and 6 km, giving good coverage
throughout that region of the spatial frequency plane.  The observations 
were made at 4798 and 8638 MHz with 
bandwidths of 100 MHz in polarization mode,
giving data on all four Stokes parameters.

For each observation, the flux calibrator source PKS B1934-638
was observed at the beginning and end of the run.  The phase calibrator 
chosen was the point source, PKS B0454-810 for the first four data sets and 
PKS B0530-727 for the fifth data set.  
In all observations, the phase calibrator was observed 
at approximately 20 minute intervals to allow for proper phase correction. 
Flux densities for the calibrators are given in Table 1.

Data reduction was carried out with the {\it MIRIAD} package (Sault, Teuben, 
\& Wright 1995).  
The data were calibrated and formed into dirty images before cleaning
and restoring with circular beams.  The circular HPBW sizes were 
$1.1 ''$ and $1.8 ''$ for the 3-cm and 6-cm images
respectively.  The images were made with ROBUST weighting (Briggs 1995)
to maintain high angular resolution but provide as much sensitivity as 
possible.

Polarization maps were also made with the {\it MIRIAD} package.  Dirty images
were made in Stokes $Q$, $U$, and $I$, cleaned, and restored using the 
circular beams mentioned above.  To reduce the noise level, each image
was convolved with a 5'' beam.  The {\it MIRIAD} task IMPOL was then used to 
combine the convolved maps in the different Stokes parameters to make 
a polarized intensity map and a position angle map.  These maps were used 
to determine the strength
and direction of the magnetic field.  The {\it MIRIAD} task IMRM was used 
to calculate the Faraday rotation measure across the SNR. 

\section{Results}

\subsection{Morphology} 

\subsubsection{Overall SNR}

The final radio images at 6 and 3 cm are shown in Figures 1a and 1b,
respectively.  The supernova remnant is 180 arcseconds
$\times$ 195 arcseconds in size, slightly elongated in roughly 
the east-west direction.
Throughout the paper, we will adopt a distance to the LMC of 50
kpc (Feast 1999).   Using this, we find a 
linear size for the remnant of 44 pc by 47 pc (north-south and east-west
respectively).  At 6 cm, the remnant has diffuse emission across its
entire face.  The eastern side is twice as bright as the western edge, with a 
steady gradient showing decreasing emission from east to west.  This 
brightness gradient can be seen in the one dimensional slices through the
6-cm image shown in Figure 2.  Although the east-west slices may
indicate the presence of a shell, they still show that the central emission
from the remnant stays well above the background.
Based solely on this morphology, the remnant may be classified as a 
filled or Crab-like or perhaps a composite remnant.

The structure of the SNR is less obvious in the 3-cm image.  
The morphology of the remnant
from this image may indicate edge-brightening on the north-eastern and 
south-western edges, similar to what is seem at 6 cm.
Unfortunately, because of the smaller beamwidth, the surface brightness of 
the remnant at 3 cm is significantly
lower than at 6 cm.  The peak brightness of the map is just over the 
3-$\sigma$ ($\sigma_{\rm 3 cm} = 3.4 \times 10^{-5}$ Jy beam$^{-1}$) 
noise level, 
in contrast to over 6-$\sigma$ ($\sigma_{\rm 6 cm} = 6.0 \times 10^{-5}$ Jy 
beam$^{-1}$) for the 6-cm image.  The fainter emission is lost in the noise 
in the 3-cm image.  
Natural weighting was tried to increase sensitivity to the large-scale
emission, but it also emphasized the side lobes and did not significantly
improve the image quality.
Therefore, nothing can be said about the morphology of the remnant's interior
at 3 cm.

Figures 3a and 3b show {\it ROSAT} HRI (Williams 1999) and H$\alpha$ 
(Magellanic Clouds Emission Line Survey; Smith 1999) images of 
N206, overlaid with 6-cm radio contours.  The {\it ROSAT} HRI image shows 
a somewhat elliptical central brightening with a gradual
and isotropic decrease in brightness with increasing radial distance
from the remnant's center.  There is no apparent brightening on the
eastern or western edges.  In contrast, the radio image does not show strong
emission from the remnant's center; the brightest region is in the
eastern half of the remnant.

The H$\alpha$ image shows a very filamentary structure
and a distinct shell to the remnant.  
These filaments and shell are also prominent in images taken in 
[\ion{S}{2}] and [\ion{O}{3}] (Lasker 1977).

Although the optical emission is primarily in filaments, there is some 
diffuse optical emission in the center of the remnant in the H$\alpha$ image.
As seen in echelle data, this central emission represents a stationary 
component (attributed to emission from the nearby \ion{H}{2} region), whereas
the emission from the expanding SNR has a velocity of around 250 km s$^{-1}$ 
(Chu \& Kennicutt 1988).  To further test the nature of this central 
emission, we interpolated between the 
integrated radio and X-ray flux densities.  We found that the estimated flux 
density from synchrotron emission in the optical is several orders of magnitude
too low to appear in the H$\alpha$ image.   
Therefore, this diffuse optical emission does indeed
seem to arise from the nearby \ion{H}{2} region rather 
than from optical synchrotron emission from the SNR.  

\subsubsection{Peculiar Linear Feature}

In the eastern side of the remnant is a linear feature, 
oriented east-west and stretching from a point roughly 25 arcseconds west
of the center of the remnant. 
This feature is approximately 55 arcseconds (13 pc) in length, ending
approximately 20 arcseconds from the eastern edge of the SNR.  
The widest point of this feature is at its western end, narrowing
steadily toward the east with a Full-Width at Half Maximum (FWHM) of less 
than 5 arcseconds at
its eastern tip.  Included in Figure 2 are two
1-D slices through the 6-cm data (slices 4 and 5).  Once Gaussian profiles are fitted to
each slice, the FWHM of the feature can be determined, along with the error
of the fit.
The results of this process are given in Figure 4,
showing the FWHM of the linear feature 
as a function of distance from the remnant's center.

This linear feature has never been detected before at any wavelength.
Although the optical image shows that the remnant has a filamentary
structure, there are no filaments that appear associated with the 
linear feature
in the radio image.  The X-ray image shows that the brightest emission
is concentrated in the remnant's center and appears to be reasonably 
spherically distributed.  In addition, there is no point source located
at either end of the feature in any of the wavelength bands.
The origin of this linear structure is not known, but will be further
examined below ($\S$ 4).

While some other SNRs exhibit long thin features, they can be seen as parts of
shell filaments visible in both radio and optical images.  One structure that
may be similar to what we see in N206 is a linear feature in G84.2$-$0.8,
but that also has a point radio source at one end (Matthews and Shaver 1980).
To our knowledge, the linear feature in N206 is the first such feature seen 
in an SNR without accompanying filamentary structure or a point source.

\subsection{Spectral Index}

Using the ATCA data,
we obtained flux densities of 0.52 $\pm$ 0.07 Jy at 4798 MHz and
0.49 $\pm$ 0.12 Jy at 8638 MHz.  There is a gradient in the background level 
due to
contamination from the \ion{H}{2} region located to the south-west of the
remnant.  Taking a mean value for the background led to the above
flux density measurements. 
Although no integrated flux density measurements in the literature cite
errors (with the exception of that by Mills et al 1984 who give an 
expected error of
around 10\%), we believe that our error at 6 cm should be comparable with the 
other measurements.

All flux density values available for N206 are listed in Table 2 and the 
spectrum of the remnant is shown in Figure 5.  The flux density scale for all
the given values, except the 843 MHz MOST result (Mills et al. 1984)
has been checked and is found to agree with the current ATNF scale based
on the calibration source PKS B1934-638 (Reynolds 1994).  The calibration 
scale used for the MOST during the early commissioning phase when the
measurement was made 
is somewhat uncertain but should be within 5\% of the current one.
A linear, unweighted regression
to the data (i.e. a power law fit) with a slope of $-$0.20 $\pm$ 0.07 is also 
plotted.  
Such a value is typical for a spectral index expected from a filled
center SNR ($S \sim \nu^{\alpha}, -0.4 < \alpha < 0$) but is flatter than is 
generally found for a shell-type remnant ($-0.8 <
\alpha <  -0.3$) (Trushkin 2000).

The faintness of the 3 cm data makes it difficult to make a spectral index
map of the remnant.  To try to increase the sensitivity, both the 3-cm and
6-cm images were convolved with $5 ''$ circular beams.  The convolved
images were both masked at the 1-$\sigma$ level and the
spectral index map shown in Figure 6 was made. 
Although the 3-cm image still suffers from low sensitivity, 
the resulting map does give reasonable results
for the linear feature since the signal to noise ratio is considerably better
in this region than in the rest of the remnant.
In the vicinity of the linear feature, the spectral index is approximately 
$-$0.2, in good agreement with the results for the entire SNR from the 
previous figure.

\subsection{Polarization}

The polarization map of the region around the linear feature in N206 is shown 
in Figure 7. The vector length represents the 6-cm polarized intensity 
and the vector direction is that of the intrinsic {\it magnetic} field after 
correcting for the Faraday rotation.
For this image, the polarized intensity at both wavelengths
was greater than 1.5-$\sigma$.
Although most of the vectors are associated with the peculiar linear 
feature,
this is a brightness effect.  The fractional polarization on the linear
feature was approximately 15\%.  Unfortunately, the signal-to-noise ratio on 
the remainder of the remnant was insufficient to reliably determine the 
magnetic field direction or the fractional polarization, although it does 
appear that there is some polarization throughout the SNR.

The rotation measure determined from the position angle
rotation between 6 and 3 cm is about
+200 radians/m$^2$, increasing slightly
as you move from west to east along the linear feature.  
After correcting for the Faraday rotation, we find that the  
magnetic field at the location of the linear feature is aligned 
nearly along the feature.  
It seems likely that this direction is representative of the field in the 
feature itself and not of the remnant as a whole.   
However, without knowledge of the global 
magnetic field orientation in the SNR, we cannot be certain
of this assessment. 

Even more interesting is the hint of a twisting of the magnetic
field near the center of the linear feature.  This can be seen as a change 
in the direction of the vectors in Figure 7 where the magnetic field vectors
are oriented roughly north-west to south-east.  There is also a variation
in the polarized intensity at this location in the center of the jet in 
the figure shown.
This spot has brighter emission in the total radio intensity
and approximately zero rotation measure.  This leads us to believe
that there could be some thermal emission at this spot from a dense clump 
with a field alignment which adds some negative Faraday rotation to bring
the mean closer to zero.
The material in the linear feature is apparently clumpy.  

\section{Discussion}

\subsection{Classification of the Remnant}

The classification of this SNR is not straight forward.  The radio emission is nearly uniform with a smooth brightness gradient from east to west across 
the face of the remnant (Fig. 2).  The X-ray emission is centrally peaked
while the optical spectral lines indicate a shell structure.  We await
XMM-Newton spectra to see if the X-ray emission is thermal or non-thermal.
The radio spectral index of $-$0.20 is typical of a Crab-like SNR and
is lower than the value expected for a shell-type remnant (Trushkin 
1999).  It may be a composite remnant in which the plerion component has 
reached the shell or the shell may just be forming.  In either case, we suggest
that there is at least a component to the SNR that may be powered by 
an unseen pulsar.  
The lack of a detectable pulsar is true for several other Crab-like SNRs, 
e.g. 3C58 (Bocchino et
al. 2001), G21.5-0.9 (Slane et al. 2000), and G328.4+0.2 (Gaensler, Dickel, 
\& Green 2000).

The more rapid decay in X-ray brightness toward the edges of the remnant
may be explained by the fact that the X-ray synchrotron emission would 
decay faster, leaving primarily radio synchrotron emission in the outer 
regions.  A pulsar could still be bright in thermal X-rays, due to its high 
temperature, but available X-ray observations have insufficient spatial 
resolution to detect a possible X-ray point source in the SNR.  The 
observed X-ray emission may be partly thermal as well.  Additional
observations are needed at higher spectral and spatial resolution. 

If the remnant is indeed Crab-like, at 47 pc in diameter, it is the largest
one known, almost twice the size of the previously largest known Crab-type 
remnant G328.4+0.2 (Gaensler, Dickel, \& Green 2000).  Following the 
method of Gaensler, Dickel, \& Green to relate the pulsar spindown energy to 
the expansion of the surrounding wind nebula, we can estimate the age of the
remnant based on the size of the apparent plerion.  By taking the wind as 
a pressure
source driving the swept-up material, one can relate the nebular age to its
size, the energy injection rate, and the ambient density.  Since a distinct
separate radio 
shell is not
observed, we will take the size of the plerion to be the entire measured
size of the remnant, 23.5 parsecs in radius.  Using the spectrum given above,
we obtain a radio luminosity ($L_{r}$) for the remnant of 
$9.25 \times 10^{34}$ 
erg s$^{-1}$ when we integrate from 100 MHz to 100 GHz.  We define $\epsilon = 
L_{r}/\dot{E} \sim (1-5) \times 10^{-4}$ (Gaensler et al. 2000) for 
Crab-type remnants, where $\dot{E}$ is the pulsar energy 
loss rate.  We will adopt the mean value, $\epsilon = 3 \times 10^{-4}$. 
Taking the initial ISM particle density prior to the supernova event of 
0.003 cm$^{-3}$, we obtain an age of 29,000 years for the remnant.  As we 
would expect, N206 is probably the oldest of the Crab-like remnants. 

This age is, of course, only a very rough estimate.  For example, the density
used may be an underestimate, especially considering
the presence of an \ion{H}{2} region southwest of the remnant.
However, increasing the density would only lead to an increased estimate
for the age.  

\subsection{Peculiar Linear Feature}

The linear feature seen in both the 3 and 6-cm images is unusual.  
It is bright and present in both the 3 and 6-cm images, and it is not 
associated with any filaments in the remnant.  
The data used to create the maps shown in Figures 1a and 1b
were taken on five separate observing nights, spaced over a three month
time period and when maps were made from each individual data set, the linear 
feature appeared in every one.  Thus, it is extremely unlikely that it could
be an instrumental problem.  Our conclusion is that it is a real feature
that requires explanation.

One question that needs to be addressed is whether the
feature is actually associated with the remnant or is the result of a
line of sight coincidence.  Recall that the data indicate a shallow
spectral index for the feature, similar to the spectral index for the
whole remnant.  If the linear feature were a jet,
as from a radio galaxy along the line of sight, one would expect to find 
a much steeper spectral index (e.g. Jarvis et al 2001).
Also, if a radio galaxy were responsible for the linear feature, we might 
expect the galaxy itself to appear in the images.  There is no evidence for
a point or extended source at either end of the linear feature in either the
radio or the optical images.  In addition, the linear feature is aligned
radially with the center of the remnant.  Such a chance alignment is 
unlikely unless the feature is physically associated with the SNR.

We conclude that this linear feature is most likely to be physically
associated with the remnant.  If indeed this is true, the mechanism
by which this emission is created must be energetic enough to account
for the observed radio emission while low enough in energy to not produce 
detectable optical or X-ray emission.   Other key constraints are:  the
lack of a point source anywhere along the linear feature, the
centrally-concentrated nature of the emission at all wavelengths, and the
wedge-like morphology of the linear feature.  

\subsubsection{Constraints on the Physical Origin of the Linear Feature}

We have considered several possible mechanisms that might give rise to the 
linear feature, and find that the available data powerfully constrain  
these scenarios.  For example, a key constraint arises from the spectral
index information and the filled radio morphology, which lead us to 
classify the remnant as Crab-like.  The synchrotron emission
from Crab-like remnants is powered at least in part by a pulsar, although 
the pulsar is often not detected.   Thus, any explanation of the remnant should
include a pulsar.  Moreover, the central X-rays and the filled-center
radio morphology seem to require that the pulsar remains near the center 
of the remnant. 

The data thus drive us to explain the linear feature in a model with a 
centrally located pulsar.
The first possibility is that the feature is a radio jet from the central
pulsar.  There are two major problems with this notion.  First, the
feature is widest near the center and is very narrow at its farthest end
(cf Figure 4),
the opposite of what would be expected from a jet emanating from a central
pulsar.  Also, the feature appears to start at approximately 6 parsecs
(25 arcseconds) 
from the center of the remnant, which is unlikely for a jet created by 
a pulsar that originated in the center of the SNR.  If it is indeed a jet, 
it must be rather energetic to be 
bright out to approximately 13 parsecs away from the emitting source. 
Such a bright emitting
source aligned in the plane of the sky should have an equally bright jet on
the opposite side; such emission is not observed. If the jet is not aligned
in the plane of the sky, the opposite jet may not be visible due to
Doppler beaming; yet there might still be emission from any material
that interacts with the relativistic particles in the jet,
which is not observed.  It is also unlikely that the
relativistic expansion of material needed for Doppler beaming would exist 
that far from the emitting source.  In addition, any inclination to the 
plane of the sky would mean that the size of the jet is greater than 
13 parsecs, and thus it would have to be more energetic.   The
more energetic the feature, the more likely it would be to appear in optical
and X-ray bands.    

Another alternative to account for the enhanced synchrotron emission
is shock excitation.  A shock wave can create a density 
enhancement in the remnant's material.  This density enhancement will 
also cause an enhancement in the magnetic field as the field lines are
compressed.  
The shape of the linear feature supports the notion that it was created by an 
object moving supersonically through the ambient interior of the SNR and 
creating shock waves.
Figure 4 shows that there is a correlation between the FWHM 
of the linear feature and its distance from the remnant's center.  If 
indeed the object originated in the center and moved toward the edge of the
remnant at greater than the sound speed, we would expect to see a feature with 
this shape.

From the opening angle of the feature, we can determine the Mach number
at which the object is travelling and compare the resulting length and 
time scales with the estimated lifetime stated previously.  
If this is a bow shock (due to motion in the plane of the sky), then the 
Mach number, M, is M $= [\sin(\theta/2)]^{-1}$, where $\theta$ is the 
full opening angle.  
From a linear regression, $\theta$ = 0.222 $\pm$ 0.013 radians, 
giving a Mach number of 9.0 $\pm$ 0.5.  If we use $ \sim 10^6$ K as
a representative temperature for the remnant, this leads to a spatial
velocity for the moving object of around 800 km/s.  (If the jet is inclined
to the plane of the sky, the Mach number and velocity increase by roughly
a factor of $(\sin i)^{-1}$, where $i$ is the inclination angle measured
relative to the line of sight.)  The eastern edge of the
feature is 80 arcseconds from the remnant's center,
corresponding to a linear distance of 19.4 pc.
This leads to an age of 23,000 years for the remnant if
the object originated in the center and maintained a nearly
constant velocity.  Given the uncertainties, this estimate is consistent
with that found above for the SNR expansion.

Since there is consistency between the shape and the likely dynamical age for 
the feature, we proceed by speculating what type of object was responsible
for its creation. 
If a pulsar is not responsible for creating the linear feature, then 
from the arguments made above, the most likely candidate is a companion 
star.  From the nearly circular shape of the remnant, we 
make the assumption
that the explosion site was very near the geometrical center of the 
radio emission.  
The inferred companion velocity of 800 km/s places strong constraints
on the binary system.  Since the kick velocity of the pulsar must be small,
the companion velocity must be due to its orbital motion.  But in this case,
the separation between the stars is very close:  $a \sim 3 R_\odot \ 
({M_1/10 \, M_\odot})$, and thus we demand a very tight binary system.
To avoid tidal destruction of the stars requires that they be relatively
dense, with a massive primary having $\rho \ga 5 \ {\rm g/cm^3}$, which is
far higher than in a normal supergiant.  We are thus forced to require that
the pre-supernova be in a dense form of a bare helium (or CO) core, and thus
we also demand a relatively dense companion -- either a low-mass main sequence
star or a compact object.

In fact, Nomoto et al.\ (1994) have proposed an evolutionary path that would
lead to such systems and suggest that these might be the progenitors of Type
Ic supernovae.  Nomoto et al.\ estimate that the frequency of such events is
about $\sim 9\%$ the rate of Type II events.  Thus, while these systems would
be rare, they would not be inordinately so.  Consequently, we speculate that
N206 could be the remnant of a binary system that led to a Type Ic supernova, 
with a low-mass main sequence or compact companion (white dwarf or neutron 
star) that has led to the peculiar linear feature.

While this model is able to explain the X-ray and radio morphologies, 
it is unfortunately not yet adequate to reproduce all 
of the observational results.  The passage of the star will produce shock 
waves analogous to those in a shell-type supernova remnant, in which the shock 
waves will
propagate through the surrounding material, causing compression of that 
material and increased synchrotron emission.  Typically, the 
brightest emission is expected from the shock front.  Thus we would expect
the production of a conical structure whose point is located at the 
position of the moving star.  The propagating shock wave marks the outer 
surface of the cone.  Since the cone would have a relatively evacuated center 
and a density enhancement behind the shock, we would expect to see an 
edge-brightened morphology, as in a shell-type supernova remnant. However,
the brightest part of the linear feature is along the central line
of the feature rather than at the edge of the wedge shape (center-brightened 
rather than edge-brightened).  Possibly the interaction of the moving star with
the supernova remnant's material causes a large portion of the particles to 
become trapped at or near the center of the conical feature, 
rather than being carried along by the shock front, or the particles are 
allowed to diffuse back into the center of the cone.  Perhaps the non-uniform 
magnetic field pattern and possible clumping found from the polarimetry data 
above can help
to diffuse the particles back into the center of the linear feature,
creating the center-brightening.  This line of reasoning requires 
further study.

\section{Conclusions}

We present high-resolution ATCA images of the LMC supernova remnant
N206 at 3 and 6-cm wavelengths.  The remnant is roughly circular and 
centrally brightened in the radio, with a possible radio shell.
Observations at X-ray wavelengths show a  
centrally-filled morphology.  
The optical observations show bright filaments around the edges,
forming a full optical shell.

We found integrated flux densities of 0.52 and 0.49 Jy at 6 and 3-cm
respectively.  These values are consistent with previous radio measurements.
Overall, the radio data indicate a spectral index $-$0.20 $\pm$ 0.07 for the 
remnant, typical of a Crab-type SNR.  

Most peculiar to this SNR is the detection of a narrow linear feature
seen in radio but
undetected at other wavelengths.  The feature
has a roughly constant opening angle of 12 degrees, stretching from
inside of the eastern edge of the remnant and becoming lost in
the noise near 25 arcseconds from the center of the remnant.  Taking
50 kpc as the distance to the LMC, the feature is 13 parcsecs
in length.

The first polarimetric maps were made for the remnant.  
We found linear polarization on the linear feature, the only region 
bright enough for reliable measurements.   The polarized intensity was 
roughly  
15\% over the linear feature where the signal to noise ratio was larger
than 3-$\sigma$.
The magnetic field seems to be aligned along the feature with a twist in 
the center.  The polarimetry map is only reliable in the vicinity
of the linear feature so a magnetic field map for the entire remnant 
could not be made.

From the observational data, we were able to place several constraints
on our proposed model for the linear feature.  From the center-filled 
radio morphology and spectrum, we conclude that a pulsar should exist in 
the SNR.  Since bright X-ray emission is generally associated with
a pulsar, the pulsar is not likely to be located at either end of the linear 
feature where little X-ray 
emission is detected.  Rather, the X-ray emission near the center of the 
remnant led us to choose a model which had the pulsar spatially coincident
with the strongest X-ray emission.

To satisfy the above constraints, our proposed model to explain the presence of
the linear feature is the passage of a low-mass stellar object through the 
remnant's material at approximately
Mach 9.0, which corresponds to 800 km/s for a medium at 10$^6$ K.  The 
resulting shock waves produce density and magnetic field
enhancement behind the object, causing the enhanced synchrotron emission.
From the lack of X-ray emission associated with the feature, we speculate
that the object responsible is a low-mass companion star to the 
supernova progenitor that was ejected from the tight binary system at the time
of the explosion.   Such a scenario has been proposed for Type Ic
supernovae.  This model is still insufficient to properly explain
all of the observations, particularly the lack of limb brightening in the 
wedge-shaped feature.

Better constraints on the proposed model can be made when higher resolution
X-ray images and X-ray spectra become available for the remnant.  The
discovery of a pulsar in the remnant may clarify the 
situation and should help to constrain the model discussed above.

\acknowledgements

We would like to thank Miroslav Filipovic for contributing the fifth data set
for our images.  We would also like to thank Rosa Williams and Chris Smith
for providing the Rosat and H$\alpha$ images.  Thanks also to Ronak Shah
for help with gaussian fitting and error analysis, to Martin Guerrero for help
with determination of the feasibility of optical detection, and to 
Ron Webbink for enlightening discussions on binary systems.  Finally, we are 
grateful to the anonymous referee for comments that led us to improve 
the paper.  JRD was 
supported, in part, by the Campus Honors Program at UIUC, and the work of BDF 
was supported by the National Science Foundation under Grant AST-0092939.

\onecolumn
\clearpage

\begin{deluxetable}{crrrrrrrrrrr}
\tabletypesize{\small}
\tablecaption{Observing Parameters and Calibrator Flux 
Densities ($S_{\nu}$ = Flux Density) \label{tbl-1}}
\tablewidth{0pt}
\tablehead{
\colhead{Date} & \colhead{Configuration}   & \colhead{Calibrator}   &
\colhead{3-cm $S_{\nu}$ (Jy)} &
\colhead{6-cm $S_{\nu}$ (Jy)}  
}
\startdata

Aug. 7, 1997 & 750B & 0454-810 & 1.657 & 1.092 \\
Aug. 27,1997 & 1.5C & 0454-810 & 2.030 & 1.266 \\
Oct. 10,1997 & 375 & 0530-727 & 0.265 & 0.25 \\
Oct. 23,1997 & 750C & 0454-810 & 2.509 & 1.744 \\
Nov. 11,1997 & 6C & 0454-810 & 2.393 & 1.829 \\
All & All & 1934-638 & 2.84 & 5.83 \\

\enddata 
\end{deluxetable} 

\clearpage

\begin{deluxetable}{ccclllllllll}
\tabletypesize{\small}
\tablecaption{Flux Densities of SNR N206 \label{tbl-1}}
\tablewidth{0pt}
\tablehead{
\colhead{Frequency} & \colhead{$S_{\nu}$}   & \colhead{Quoted}   &
\colhead{Source of Plotted} &
\colhead{Reference} \\[-0.05in] 
\colhead{(MHz)} & \colhead{(Jy)} & \colhead{Uncertainty} & \colhead{Uncertainty} 
& \colhead{for $S_{\nu}$}
}

\startdata

408 & 0.7 & none given & background removal & Matthewson \& \\[-0.1in]
   &    &  & estimates by others &  Clarke 1973 \\
843 & 0.591 & 10\% & rms noise & Mills et al. 1984 \\
4798 & 0.52 & $\pm$0.07 & background subtraction & this paper \\
8638 & 0.49 & $\pm$0.12 & background subtraction & this paper \\
14700 & 0.27 & $\pm$0.22 & background subtraction & Milne et al. 1980 \\[-0.1in]
      &      &           &  for peak   &   \\

\enddata 
\end{deluxetable}

\clearpage

\begin{figure}
\epsscale{0.7}
\plotone{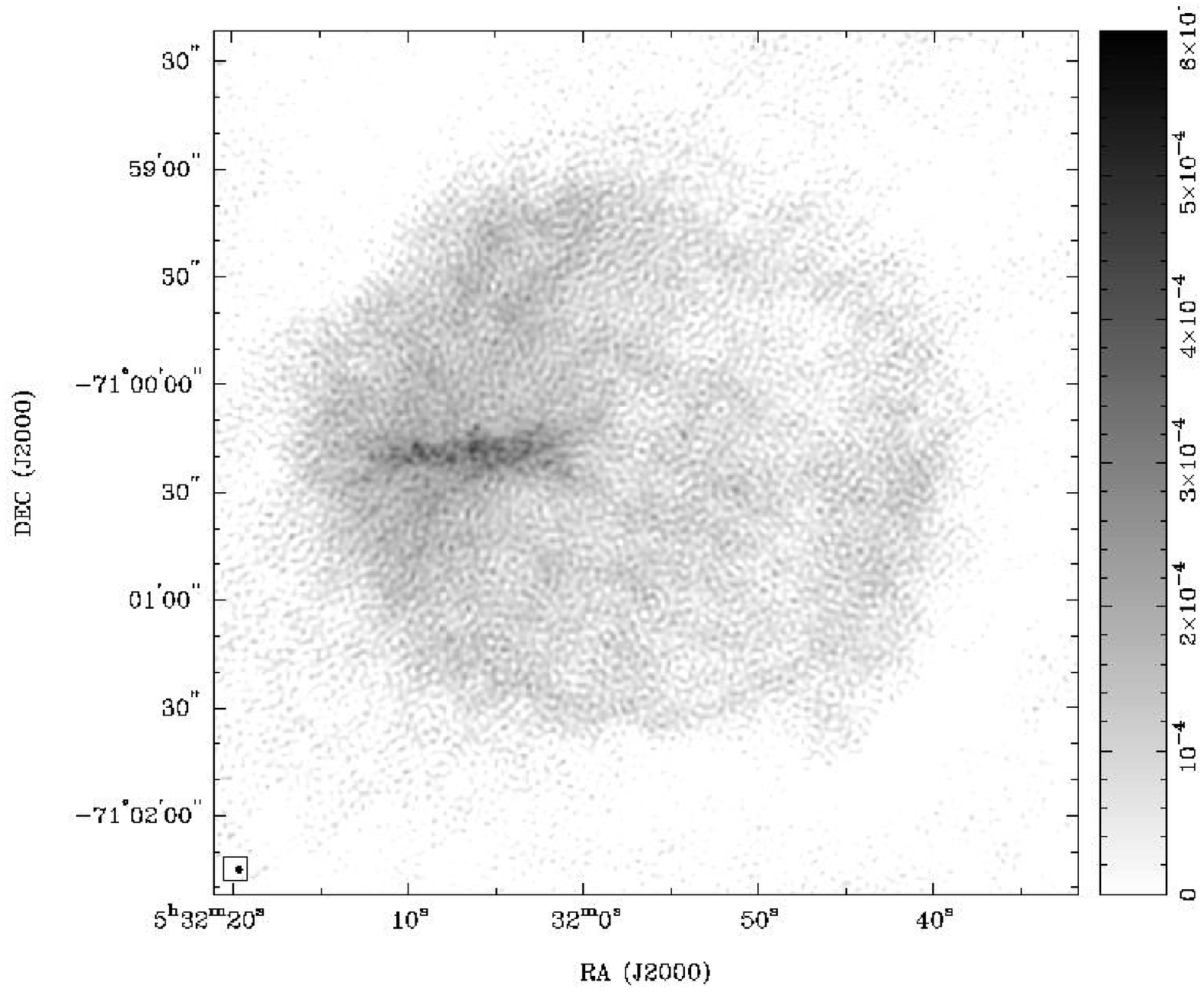}
\epsscale{0.7}
\plotone{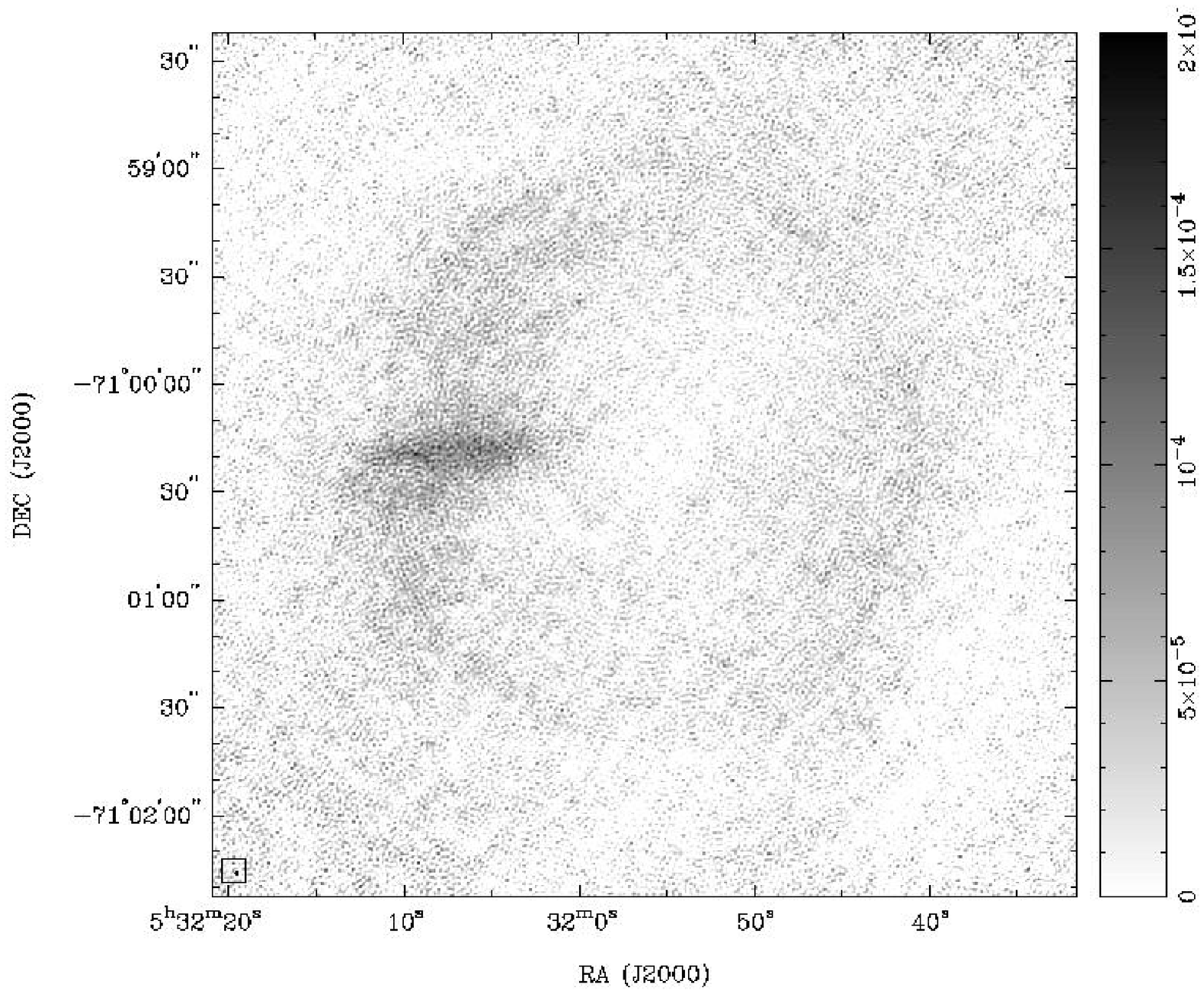}
\caption{6-cm (upper panel) and 3-cm (lower panel) radio images of the SNR 
N206.  The units on the wedges are Jy beam$^{-1}$.  Circular beams were used 
with beamwidths of $1.8 ''$ and $1.1 ''$ for the 6-cm and 3-cm images, 
respectively.  Beam sizes are given in the lower left-hand corner of 
each image.
}
\end{figure}

\clearpage 

\begin{figure}
\plotone{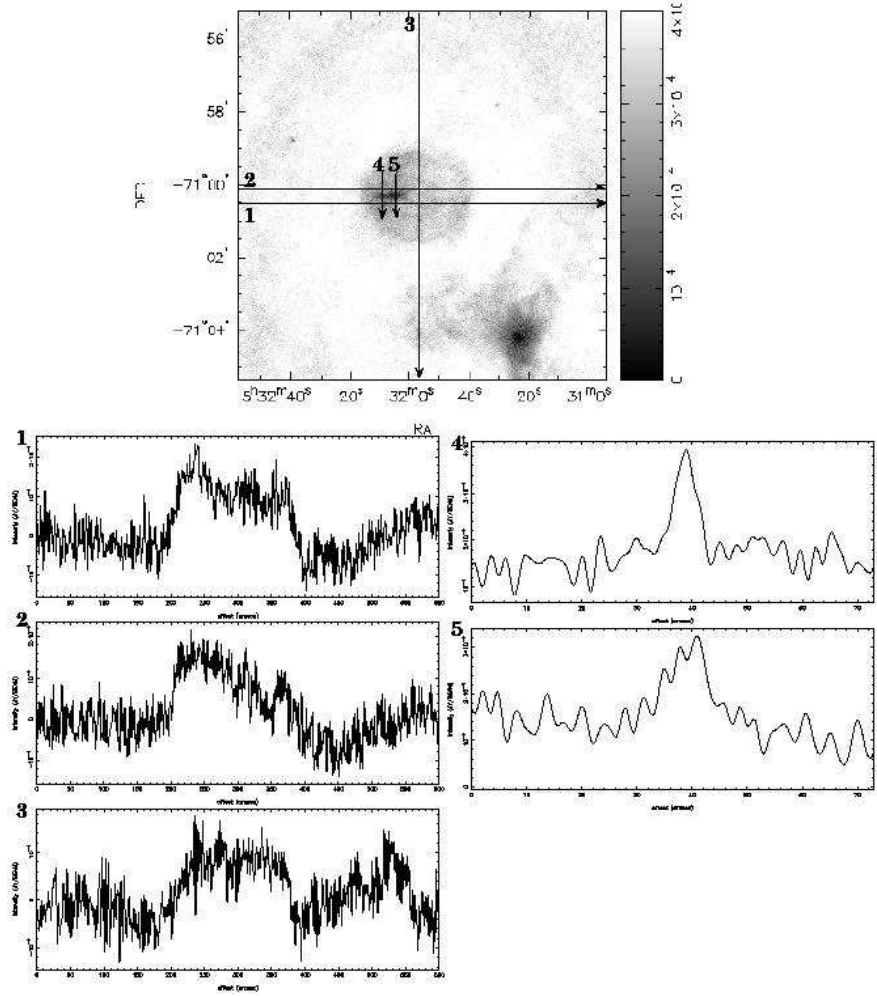}
\caption{1-dimensional slices through the 6-cm image.  
The bright feature at the lower right in the 6-cm image is the
\ion{H}{2} region part of N206.
}
\end{figure}

\clearpage

\begin{figure}
\epsscale{0.7}
\plotone{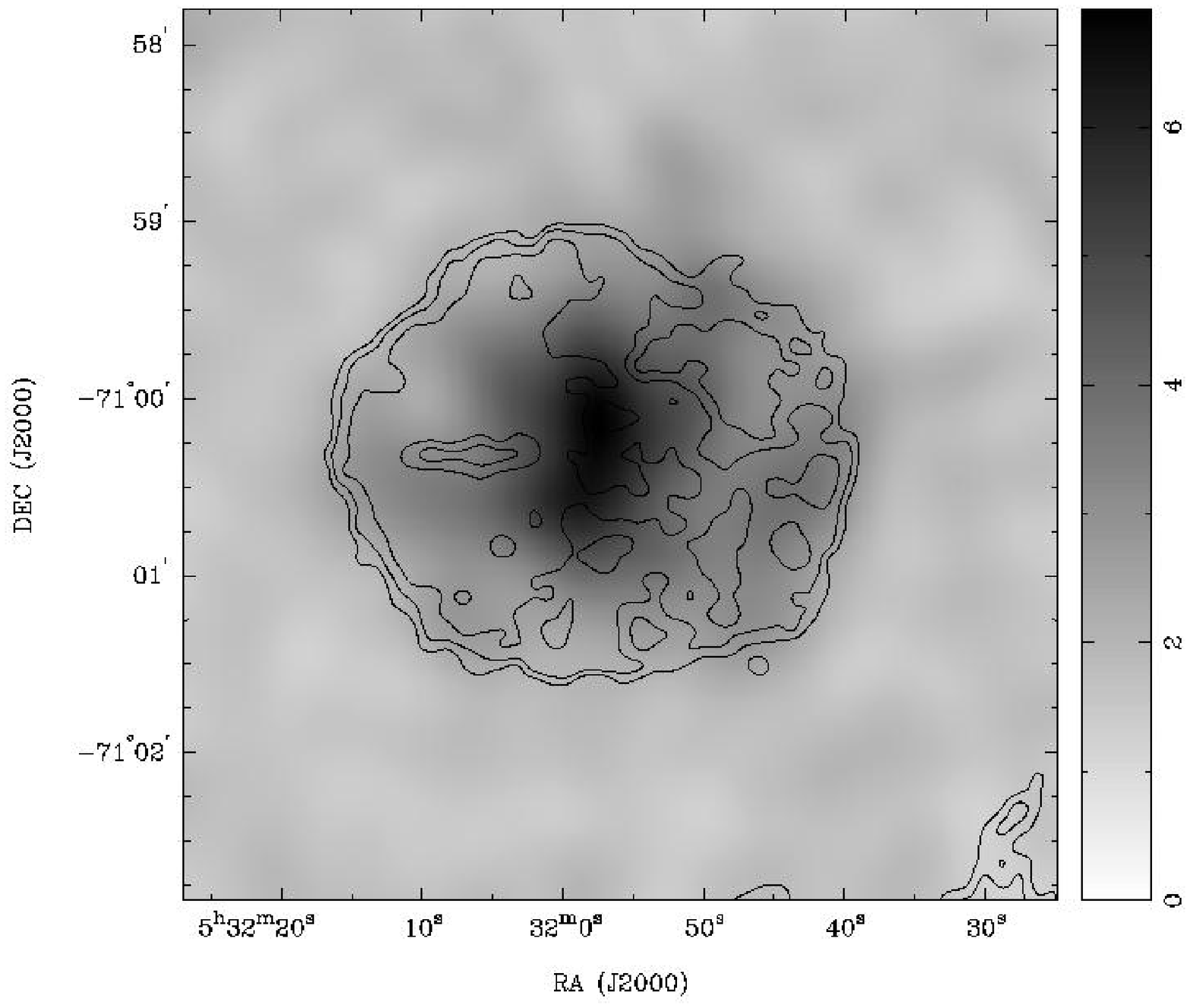}
\epsscale{0.7}
\plotone{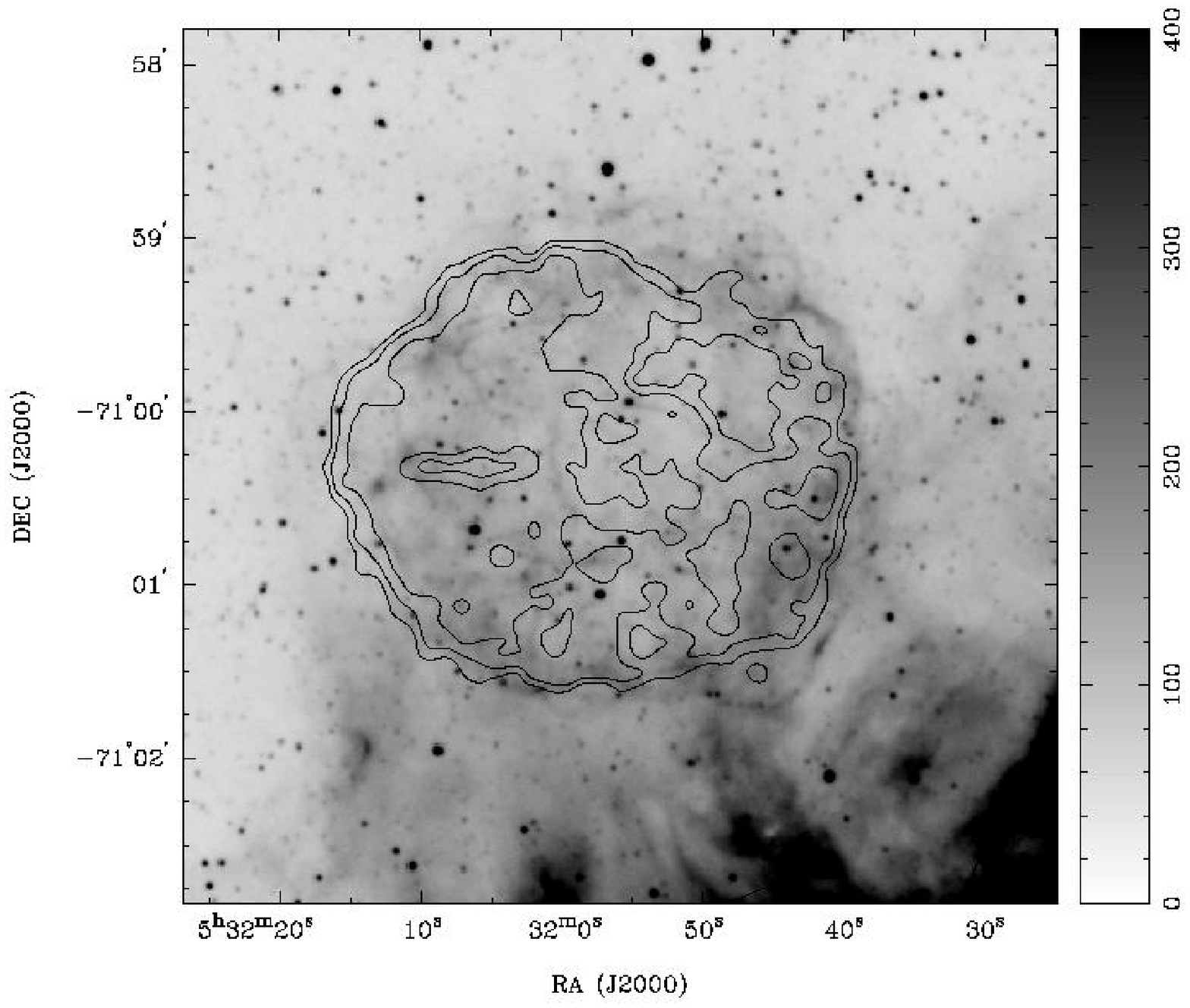}
\caption{{\it ROSAT} HRI (upper panel) and H$\alpha$ (lower panel) images 
with 6-cm radio contours.  The grayscales in the upper and lower figures
show the X-ray and optical counts, respectively.  The contours shown in both 
images are 90, 75, 50, 35, and 25\% of the peak radio intensity.
}
\end{figure}

\clearpage

\begin{figure}
\plotone{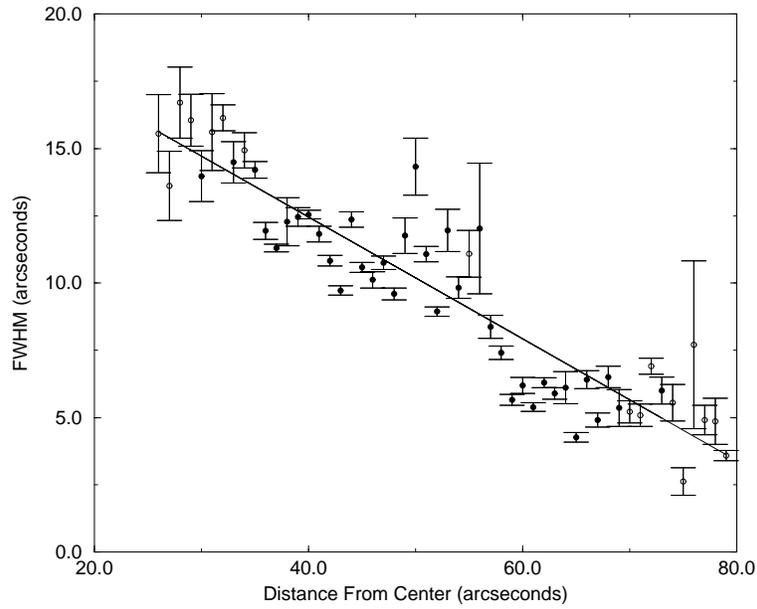}
\caption{FWHM of the linear 
feature as a function of distance from the center:  The regression
slope = $-$0.226 $\pm$ 0.013 and the opening angle, $\theta$ = 12.7$^\circ$.  
The error 
bars represent uncertainty in the individual Gaussian fits to the 1-D slice 
data taken across the linear feature.  Closed circles denote slices where 
the peak intensity of the slice was greater then 3 $\sigma$ above the 
brightness of the surrounding remnant.  Open circles represent slices for 
which the peak intensity of the slice was between 2 and 3 $\sigma$ above the
brightness of the surrounding remnant.   
}
\end{figure}

\clearpage

\begin{figure}
\plotone{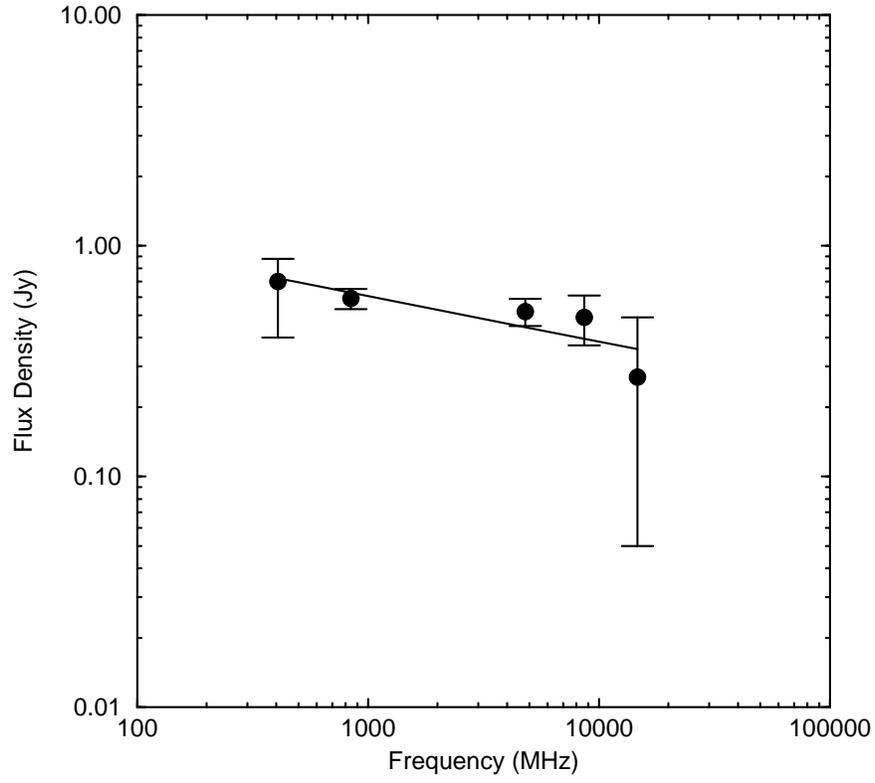}
\caption{Radio spectrum of the SNR N206.  Flux density values and 
error estimates are given in Table 2.
}
\end{figure}

\clearpage

\begin{figure}
\plotone{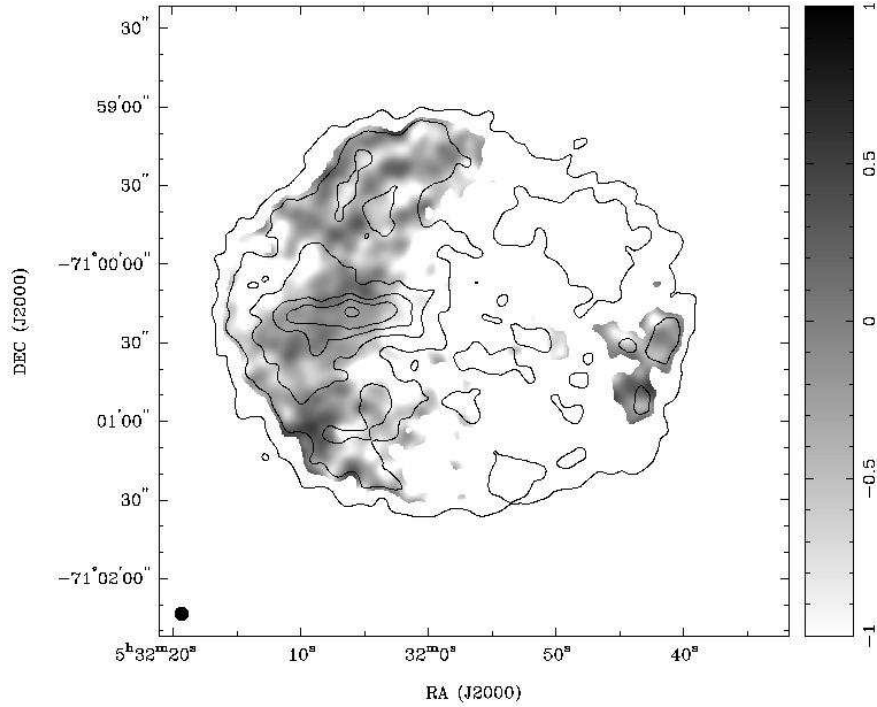}
\caption{Spectral index map of the SNR N206.  The grayscale denotes
the spectral index across the remnant made from 3 and 6-cm images convolved
to 5 ''.  
The contours denote 6-cm total intensity at 95, 75, 60, 45, 
30, and 15 percent of the peak intensity (5.694 $\times$ 10$^{-4}$ Jy 
beam$^{-1}$).  The 6-cm total intensity was also convolved to 5 ''.  The 
beam size is shown in the lower left.
}
\end{figure}

\clearpage

\begin{figure}
\plotone{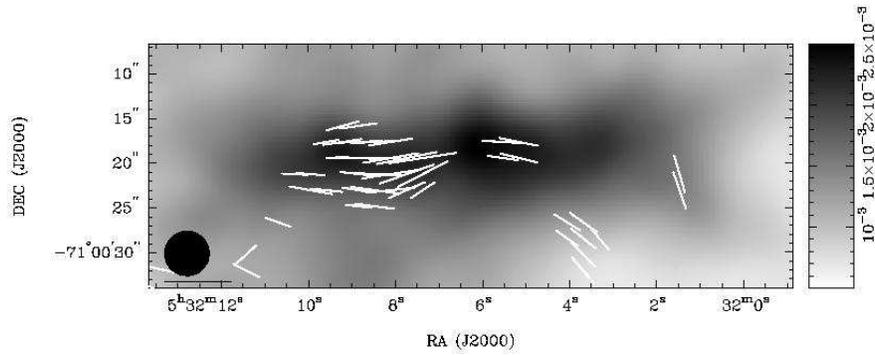}
\caption{Polarization Map of the Linear Feature in N206 at 6 cm:  The lengths of the vectors 
are scaled to 
the 6-cm polarized intensity.  The longest vector corresponds to a 
polarized surface 
brightness of 6 $\times$ 10$^{-5}$ Jy beam$^{-1}$, the length denoted
by the bar at the lower left.  The orientation of the 
vectors denotes the direction of the magnetic field.  The grayscale shows the
total intensity in Jy beam$^{-1}$. The beams were convolved to 5 ''.
The beam width is shown
in the lower left corner.}
\end{figure}

\end{document}